\documentclass{aa}  

\usepackage{graphicx}
\usepackage{txfonts}
\usepackage{lipsum}
\usepackage{subcaption}         
\usepackage{lscape}             
\usepackage{placeins}           
\usepackage[colorlinks=true, linkcolor=blue, citecolor=blue, urlcolor=blue]{hyperref}

\begin{document}

   \title{When dust growth takes over: the origin of dust in post-reionization galaxies}

\authorrunning{O. Osman et al.}

   \author{Omima Osman\inst{1, 2}\fnmsep\thanks{E-mail: omima.osman@inaf.it}, 
   Gabriella De Lucia\inst{1, 3}, Andreas. L. Faisst\inst{4}, Lizhi Xie\inst{5}, Fabio Fontanot\inst{1, 3}, Michaela Hirschmann\inst{1, 6}}

      \institute{INAF -- Astronomical Observatory of Trieste, via G.B. Tiepolo 11, I-34143 Trieste, Italy
      \and  University of Khartoum -- Department of Physics, Al--Gamaa Ave, 11115 Khartoum, Sudan
      \and IFPU -- Institute for Fundamental Physics of the Universe, via Beirut 2, I-34151 Trieste, Italy
      \and Caltech/IPAC, 1200 E. California Boulevard, Pasadena, CA 91125, USA
      \and Tianjin Astrophysics Center, Tianjin Normal University, Binshuixidao 393, 300384 Tianjin, China
      \and Institute for Physics, Laboratory for Galaxy Evolution, EPFL, Observatoire de Sauverny, Chemin Pegasi 51, 1290 Versoix, Switzerland
}

   \date{Received September xx, 2026}

 \abstract{The post-reionization epoch ($z\sim4$–6) marks a critical phase in galaxy evolution, during which galaxies rapidly assemble their stellar mass, enrich their interstellar medium, and build up substantial dust reservoirs. The physical processes responsible for such rapid dust enrichment remain uncertain. We investigate the dust enrichment history of the ALPINE–CRISTAL–JWST (ACJ) sample using the semi-analytic galaxy formation model GAlaxy Evolution and Assembly (GAEA), which follows galaxy assembly, chemical enrichment, and dust evolution. We select model galaxies matching the observed stellar masses, star formation rates, and metallicities of the ACJ sample, compare their dust properties with observed scaling relations, and trace their enrichment histories and environments. The model reproduces the observed dust mass–stellar mass relation and the dust-to-stellar mass, dust-to-gas, and dust-to-metal ratios as functions of stellar mass and metallicity. Reproducing these constraints requires efficient dust growth in the interstellar medium; stellar dust production alone cannot account for the observed dust reservoirs, even under the most optimistic assumptions. The combination of stellar mass, star formation rate, and metallicity provides a good predictor of galaxy dust content. Our model suggests rapid dust build-up consistent with the previously noted rapid metal enrichment, with evidence for a transition around $z\sim8$–9 below which dust growth becomes the dominant dust formation channel. Finally, ACJ-like galaxies preferentially reside in locally overdense environments, suggesting that environment may have contributed to their accelerated evolution. Taken together, these results support rapid and coupled stellar, metal, and dust build-up in ACJ galaxies during the post-reionization era.}



\maketitle
\nolinenumbers

\section{Introduction}

The post-reionization epoch (post-EoR; $z \sim 4$--6) marks a critical transition from primordial to more mature phases of galaxy evolution (e.g. \citealp{Kauffmann99}; \citealp{Riechers14}; \citealp{daCunha15}; \citealp{Faisst22}; \citealp{Herrera25}; \citealp{Faisst26}). During this period, galaxies rapidly build up their stellar mass and enrich their interstellar medium through intense star formation activity, which, globally, continues to rise and reaches its peak at cosmic noon ($z \sim 2$--3). The physical drivers of this accelerated evolution are, however, still under debate. In addition to the internal processes governing gas accretion, star formation, and chemical enrichment, the surrounding environment may also play an important role. Galaxies residing in overdense regions are expected to assemble earlier, experience enhanced gas accretion and more frequent interactions, leading to more rapid growth than galaxies in average-density environments (e.g. \citealp{Overzier16}; \citealp{Alberts22}). Recent observations suggest that at least a subset of massive star-forming galaxies in the post-EoR era may preferentially reside in such environments (\citealp{Lemaux18}; \citealp{Ginolfi20}; \citealp{Loiacono21}; \citealp{Romano21}), although the role of environment in shaping their rapid evolution remains poorly understood (\citealp{DeLucia26b}).

Another defining characteristic of galaxies during this epoch is the presence of massive dust reservoirs, with dust masses reaching up to $10^9$ M$_\odot$ (e.g. \citealp{Rowlands14}; \citealp{daCunha15}; \citealp{Dudzeviciute20}; \citealp{Pozzi21}). The channels responsible for such rapid dust enrichment on short timescales (less than $\sim1$ Gyr is available to build the observed dust reservoirs) remain under debate (e.g. \citealp{Popping17}; \citealp{Pozzi21}; \citealp{Nanni25}; \citealp{Algera25}). At these early epochs, galaxies are expected to have relatively low metallicities, which limits the efficiency of the dominant dust formation channel identified in the local Universe, namely dust growth in the ISM (e.g. \citealp{Remy14}; \citealp{Grossi15}; \citealp{Galliano21}). At the same time, Asymptotic
Giant Branch (AGB) stars may not have sufficient time to significantly enrich the ISM with dust (but see \citealp{Valiante09} for a counter-view), which leaves supernovae as the likely main source of dust (e.g. \citealp{Michalowski15}; \citealp{Ferrara16}; \citealp{Schneider24}).

Over the past decades, a large amount of multiwavelength observations have been accumulated decoding galaxy properties. Such observations provide measurements of stellar mass, gas mass, star formation rate (SFR), metallicity, and dust content, enabling stringent constraints on theoretical models. In particular, these observations allow models to be tested against a range of scaling relations, including dust mass as a function of stellar mass and SFR, as well as the dust-to-stellar mass, dust-to-gas, and dust-to-metal ratios as functions of stellar mass and metallicity (\citealp{Lisenfeld98}; \citealp{Dunne03}; \citealp{Vlahakis05}; \citealp{Eales09}; \citealp{daCunha10}; \citealp{Corbelli12}; \citealp{Santini14}; \citealp{Remy14}; \citealp{Beeston18}; \citealp{Nersesian19}; \citealp{Pozzi21}; \citealp{Algera25}). Each of these relations encodes different aspects of the interplay between dust and galaxy evolution (e.g. \citealp{Hirashita99}; \citealp{Inoue03}; \citealp{Kuo12}; \citealp{Hirashita13}; \citealp{Asano13a}; \citealp{Remy14}), and together they provide a powerful framework for understanding the origin and evolution of dust across cosmic time.

Several studies use dust scaling relations to test theoretical models that incorporate different aspects of dust evolution physics across cosmic time (e.g. \citealp{Bekki13}; \citealp{Hirashita15}; \citealp{Feldmann15}; \citealp{McKinnon16}; \citealp{Aoyama17}; \citealp{Hou17}; \citealp{Popping17}; \citealp{Gjergo18}; \citealp{Vijayan19}; \citealp{Li19}; \citealp{Triani20}; \citealp{Granato21}; \citealp{Parente23}; \citealp{Yates24}; \citealp{Osman25}). However, these models have generally been tested against heterogeneous observational datasets that often lack measurements of all the key physical properties for the individual galaxies. At high redshift, the available samples are also relatively small. Consequently, the observational constraints on dust evolution models remain limited. Recent multiwavelength surveys are now changing this picture by providing increasingly complete measurements of the stellar, gas, metal, and dust content of galaxies in the post-EoR era (e.g. \citealp{Fujimoto25}; \citealp{Burgarella25}; \citealp{Faisst26}).

In this work, we focus on the ALPINE–CRISTAL–JWST (ACJ) sample (\citealp{Fujimoto25}; \citealp{Faisst26}). ACJ represents a benchmark dataset for studying galaxy evolution in the post-EoR era and linking these systems to galaxies at both earlier and later cosmic times (\citealp{Faisst26}). It comprises a subsample of the ALPINE survey (\citealp{LeFevre20}; \citealp{Faisst20}; \citealp{Bethermin20}) for which all the key physical properties relevant to dust evolution—stellar mass, gas mass, SFR, metallicity, and dust content—are available for the same galaxies (a detailed description of both samples is provided in Section~\ref{data}). Interestingly, nearly half of the ACJ galaxies are classified as mergers or interacting systems, and several have serendipitously detected companions, suggesting that at least a subset resides in locally overdense environments. Using this unique dataset will enable us to (i) test our dust evolution model against multiple observational constraints, (ii) investigate the dominant dust formation processes during this early phase of galaxy evolution, and (iii) examine the environments of ACJ galaxies and discuss their possible role in the rapid build-up of stellar mass, metals, and dust.

The rest of the paper is organised as follows. Section \ref{sec:sec2} describes our dust evolution model within the framework of the GAlaxy Evolution and Assembly semi-analytic model (GAEA; \citealp{DeLucia24}; \citealp{Osman25})\footnote{\url{https://sites.google.com/inaf.it/gaea/home}}. The ACJ sample and its parent samples are presented in Section \ref{data}. Sections \ref{results} and \ref{Discussion} present our results and provide a discussion of our conclusion, respectively. Our findings are summarized in Section \ref{sum}.


\section{The model}
\label{sec:sec2}

For this work, we use the most recent version of the GAEA model (\citealp{Osman25}; hereafter O26). The model self-consistently implements dust formation and evolution together with chemical enrichment, and follows the dust mass in the different reservoirs as part of the baryon cycle (i.e. the stars, and the cold, hot, and ejected gas). The model assumes a Chabrier IMF (\citealp{Chabrier03}) and explicitly follows the evolution of eight metals in both the gas and dust phases (Mg, Si, O, C, Fe, S, Ca, Ti). We run the model on the Millennium (\citealp{Springel05}, MS) and Planck Millennium (\citealp{Baugh19}, PMS) simulations. MS (PMS) models a box with a side length of 500 (542.16) Mpc h$^{-1}$ and assumes the following cosmological parameters: $\Omega_\Lambda$ = 0.75 (0.693), $\Omega_m$ = 0.25 (0.307), $\Omega_b$ = 0.045 (0.04825), $n$ = 1 (0.9611), $\sigma_8$ = 0.9 (0.8288), and $h$ = 0.73 (0.68). The MS contains 2160$^3$ particles and achieves a resolution of 8.61$\times$10$^8$ M$_\odot$, while PMS contains 5040$^3$ particles and achieves a resolution of 1$\times$10$^7$ M$_\odot$. In the following, we briefly describe the model and refer the reader to the original paper (O26) for further details.

The reference physical model is a significantly modified version of the original model by \cite{DeLucia07}. \cite{DeLucia24} combine these modifications into the latest reference version. In particular, this includes a non-instantaneous recycling scheme for gas, energy, and metals (\citealp{DeLucia14}), an updated parametrisation of stellar feedback (\citealp{Hirschmann16}), a partition of the cold gas reservoir in its atomic and molecular components (\citealp{Xie17}), a gradual stripping of the hot and cold gas in satellite galaxies (\citealp{Xie20}), and an improved model for cold gas accretion onto black holes together with an explicit modelling of quasar-driven winds (\citealp{Fontanot20}). This version is calibrated to reproduce the stellar mass function up to redshift $\sim$ 3, the HI and H$_2$ mass functions in the local Universe, and the AGN bolometric luminosity function up to redshift $\sim$ 4 (see, e.g., \citealp{DeLucia24}). No further calibration was required when the dust model was implemented.

The initial source of dust is stellar dust formed in the ejecta of asymptotic giant branch (AGB) stars and Type Ia and Type II supernovae (SNe) (\citealp{Dwek98}). Once dust grains are injected into the ISM, they are subject to different destructive processes, including sputtering, sublimation, and complete or partial evaporation, driven either by supernova shocks propagating through the ISM or by interactions with hot gas (e.g. \citealp{Mckee89}; \citealp{Jones94}; \citealp{Jones04}; \citealp{Hirashita13}). Only 10–20\% of the produced grains are expected to survive (\citealp{Mckee89}; \citealp{Dwek98}), which is insufficient to account for the dust masses observed in the local and high-redshift Universe (\citealp{Jones94}; \citealp{Bertoldi03}; \citealp{Mattsson11}; \citealp{Kuo12}; \citealp{McKinnon16}; \citealp{Ginolfi18}; \citealp{McKinnon18}; \citealp{Shivaei24}). Surviving dust grains are expected to grow in size and mass within dense regions of the ISM through the accretion of gas-phase metals (\citealp{Dwek80}). This accretion process is highly efficient and can compensate for destruction processes. Depletion studies in the Milky Way are consistent with this expectation (\citealp{Duley78}; \citealp{Jones85}; \citealp{Savage96}; \citealp{Zhukovska16}; \citealp{Zhukovska18}). In DUSTY-GAEA, we explicitly follow all of the processes mentioned, linking them to the properties of galaxies and of their ISM.

Stellar dust production is implemented following the prescriptions of \cite{Dwek98}. These assume that a fraction of the produced metals (Mg, Si, O, C, Fe, S, Ca, Ti) condenses into dust grains according to a certain condensation function. The condensation function varies depending on whether the stellar source is an AGB star, SNIa, or SNII, and on the condensing element. The main parameters in these prescriptions are the condensation efficiencies.

Dust destruction by SN shocks is a complex process that requires knowledge of: (i) the energy and evolution of the exploding SN, (ii) the properties of the ambient ISM, and (iii) the properties of the dust grains. Furthermore, ISM regions are often affected by multiple SN explosions, and one should account for overlapping events. We follow the simplified approach developed by \cite{Mckee89}, which takes into account the SN energy, shock velocity, density of the ambient ISM, and SN clustering. Dust destruction through interactions with the hot gas in the hot and ejected reservoirs is implemented separately following the prescriptions of \citet{Tsai95}.

For dust growth in the dense ISM, we follow the prescriptions by \cite{Zhukovska08}. These  account for the physical properties of the ISM, the cycling of gas between different phases, and the composition of dust grains. Following \cite{Zhukovska08} and using the same parameter values adopted in this work, we consider carbonaceous grains, silicates (pyroxene and olivine), metallic iron, and silicon carbide.

In this work, we adopt the same parameter values as in O26. For completeness, these values are listed in Table 1.

\begin{table*}
\centering
\caption{Summary of the dust-physics prescriptions adopted in our semi-analytical model.}
\renewcommand{\arraystretch}{1.5}
\begin{tabular}{lll}
\hline
{Physical processes}
& {Gas reservoir}
& {Parameters} \\
\hline

{Dust formation} & Cold gas & $\delta^{SNIa-SNII}$ = 0.15, $\delta^{AGB}$ = 0.2\\
\hline
{Dust destruction} & Cold gas & $\xi_{SN}$ = 0.36, $\zeta_{SN}$ = 0.34 \\
\hline
{Dust sputtering} &  Hot and ejected gas & $\bar{a}$ = 0.1 \\
\hline
{Dust growth} & Cold gas & $\tau_{exch}$ = 20 Myrs \\
\hline
\end{tabular}\\

\tablefoot{$\delta^{\rm SNIa-SNII}$ and $\delta^{\rm AGB}$ are the dust condensation efficiencies in the ejecta of SNIa-SNII and AGB stars, respectively. $\xi_{\rm SN}$ and $\zeta_{\rm SN}$ are the fraction of clustered supernovae and the destruction efficiency of a single supernova, respectively. $\bar{a}$ is the average grain size, and $\tau_{\rm exch}$ is the timescale for gas exchange between different ISM phases, effectively the cloud lifetime.}

\label{table1}
\end{table*}
\section{The ALPINE--CRISTAL--JWST sample}
\label{data}

The ACJ sample (\citealp{Faisst26}; \citealp{Fujimoto25}) is a subsample drawn from the ALPINE sample (118 galaxies; \citealp{LeFevre20}; \citealp{Faisst20}; \citealp{Bethermin20}) with high-resolution ALMA follow-up from the CRISTAL program (\citealp{HerreraCamus25}). The ALPINE parent sample consists of galaxies with accurate spectroscopic redshifts obtained from Keck/DEIMOS and VLT/VIMOS observations (\citealp{LeFevre13}; \citealp{Hasinger18}; \citealp{Khostovan26}) in the COSMOS (\citealp{Scoville07}) and GOODS-S (\citealp{Giavalisco04}) fields. Galaxies were selected to have UV luminosities above $0.6\,L_\star$ (corresponding to $M_{\rm UV} < -20.2$) and to exclude type-I AGN based on X-Ray data. This luminosity threshold corresponds to star formation rates $\gtrsim 10\,M_\odot\,{\rm yr^{-1}}$, implying expected [C\,II] luminosities $L_{\rm [C\,II]} \gtrsim 1.2\times10^8\,L_\odot$ based on the \cite{DeLooze14} relation (see also \citealp{Schaerer20}). Targets were further prioritized using the empirical $M_{\rm UV}$--$L_{\rm [C\,II]}$ relation derived from the pilot sample of \cite{Capak15}. The ALPINE program observed [C\,II] and the rest-frame $\sim 158\ \mu{\rm m}$ dust continuum with ALMA Band~7, achieving a [C\,II] detection rate of $\sim64\%$ and a dust continuum detection rate of $\sim 20\%$.

A subset of ALPINE galaxies was subsequently re-observed at higher spatial resolution ($\sim 0.3''$) in the CRISTAL program (\citealp{HerreraCamus25}), targeting systems with stellar masses $M_\star \gtrsim 10^{9.5}\,M_\odot$, within $\pm 0.5$ dex of the star-forming main sequence (\citealp{Speagle14}), and detected in rest-frame UV HST imaging. These criteria favor massive galaxies suitable for spatially resolved ALMA observations. The ACJ sample has been built by further requiring robust ($>5\sigma$) [C\,II] detections, deep JWST/NIRCam imaging in at least four bands between $1\text{--}5\ \mu{\rm m}$, HST coverage including ACS/F814W and a redder band, and detectability of strong rest-frame optical emission lines (e.g., H$\beta$, [O\,II], [O\,III], H$\alpha$, [N\,II], [S\,II]) with the JWST/NIRSpec IFU in $\lesssim 2\,{\rm hr}$ of exposure time (\citealp{Faisst26}). High-resolution ALMA observations comparable to the NIRSpec IFU spatial scale were also required, selecting galaxies observed in CRISTAL or similar programs (\citealp{Devereaux24}; \citealp{Bethermin23}). Consequently, the final sample of 18 galaxies represents massive, UV-bright, main-sequence star-forming galaxies at $z \approx 4.5\text{--}5.7$. While not representative of the full galaxy population, excluding low-mass, faint, and heavily dust-obscured systems, the sample provides a uniquely powerful, spatially resolved, multi-wavelength view of stars, gas, and dust in typical star forming galaxies at early cosmic times.

Nine galaxies in the sample are classified as merger pairs (\citealp{Romano21}) or likely merging/interacting systems (\citealp{Posses25}; \citealp{Faisst26}). Seven are identified as AGN candidates (\citealp{Ren25}), and six galaxies are classified as both (i.e. mergers and AGN host candidates). In total, approximately $33\text{--}50\%$ of the sample shows evidence for mergers, AGN activity, or both. Despite this, the galaxies generally lie on or close to the star-forming main sequence, suggesting that the proximity of companions has yet to influence the galaxy's star formation activity. However, companions that are unresolved in the optical or near-infrared bands could bias the stellar mass estimates of some galaxies toward higher values (\citealp{Romano21}). 

\begin{figure*}
\centering
   \makebox[\textwidth][c]{\includegraphics[width=1.1\textwidth]{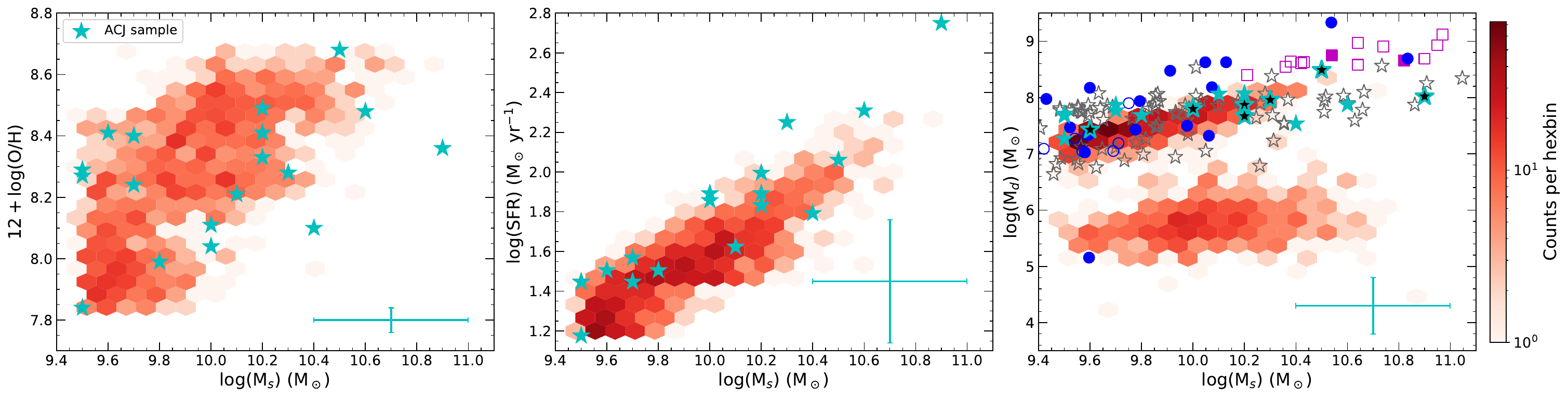}}\par    
    \caption{Gas-phase metallicity (left), SFR (middle), and dust mass (right) as functions of stellar mass in the redshift interval $4.5 \leq z \leq 5.7$. The red hexbins show the two-dimensional density distribution of the model predictions. Cyan and gray stars correspond to the ACJ and ALPINE samples, respectively. Cyan error bars indicate the typical uncertainties of the ACJ measurements.  In the right panel, cyan stars with black centres represent detections, while the ones without centres represent non-detecations (upper limits). Filled (CEERS; \citealp{Burgarella25}) and open (REBELS; \citealp{Algera25}) circles represent measurements of normal star-forming galaxies at $z \sim 4.5$ -- 6. Filled (\citealp{Dudzeviciute20}) and open (\citealp{daCunha15}) squares represent measurements of submillimeter galaxies at $z \sim 4$ -- 6.}
\vspace{-0.3cm}
\label{Z-SFR-Md}
\end{figure*}

We adopt stellar masses, star formation rates, and metallicities from \cite{Faisst25}, dust masses from \cite{Pozzi21} and \cite{Burgarella22}, and [C\,II] luminosities from \cite{Bethermin20}. \cite{Pozzi21} follow the commonly used approach of estimating dust masses using a single modified blackbody (MBB) model under the assumption of optically thin emission. In this framework, the degeneracy between dust mass and temperature cannot be broken when only a single continuum measurement is available. They therefore derive dust masses assuming two representative dust temperatures, $T_{\rm dust}=25$ and $35\,{\rm K}$, with a fixed emissivity index. We adopt the values estimated using an assumed dust temperature of 35 K, as this value is more consistent with the dust temperatures expected for galaxies at these redshifts (F. Pozzi, priv. comm.; \citealp{Faisst20}). \cite{Burgarella22} constructed an infrared composite template using 27 galaxies detected in the dust continuum with ALMA at redshifts $4 \lesssim z \lesssim 6$, 20 of which belong to the ALPINE sample. They verified that the template is not dominated by outliers. This was further supported by \cite{Sawant25} (see their Appendix~B). The resulting template was then applied to the full sample, including both continuum detections and non-detections, to derive dust properties. The fitting procedure was designed to minimize systematic uncertainties and reduce degeneracies among dust parameters.

To estimate the gas mass in the sample, we follow \cite{Dessauges20}, who infer molecular gas masses from the [CII] luminosity. We further assume that the cold gas reservoirs of high-redshift star-forming galaxies are dominated by molecular gas (e.g. \citealp{Dunne11}; \citealp{Combes18}; \citealp{Tacconi20}). The use of [CII] luminosity as a molecular gas tracer was proposed by \cite{Zanella}, who found a tight empirical correlation between [CII] emission and molecular gas mass derived from CO observations. They demonstrated that this relation holds for main-sequence galaxies out to redshift $z\sim6$, although with increased scatter at low metallicities ($12+\log({\rm O/H}) \lesssim 8.4$). The authors also discussed several caveats concerning the origin of [CII] emission, which can arise from ionized and neutral phases in addition to molecular gas (see Section~3.1 of \citealp{Dessauges20}). For a subsample of non-merging ALPINE galaxies, \cite{Dessauges20} showed that molecular gas masses estimated from [CII] luminosity are consistent with independent estimates based on dynamical masses and rest-frame 850 $\mu{\rm m}$ luminosities extrapolated from the rest-frame 158 $\mu{\rm m}$ continuum. 

Finally, following \cite{Popping17}, we estimate the total metal mass from the gas-phase oxygen abundance and using the gas mass. The metallicity, expressed as $12+\log({\rm O/H})$, is converted to a metal mass fraction assuming solar abundance ratios, and multiplied by the gas mass to obtain the total metal content. We compute the metal mass in our model using the same approach to improve consistency between observations and simulations.

\section{Results}
\label{results}
\begin{figure}
    \includegraphics[height=12cm,width=8cm]{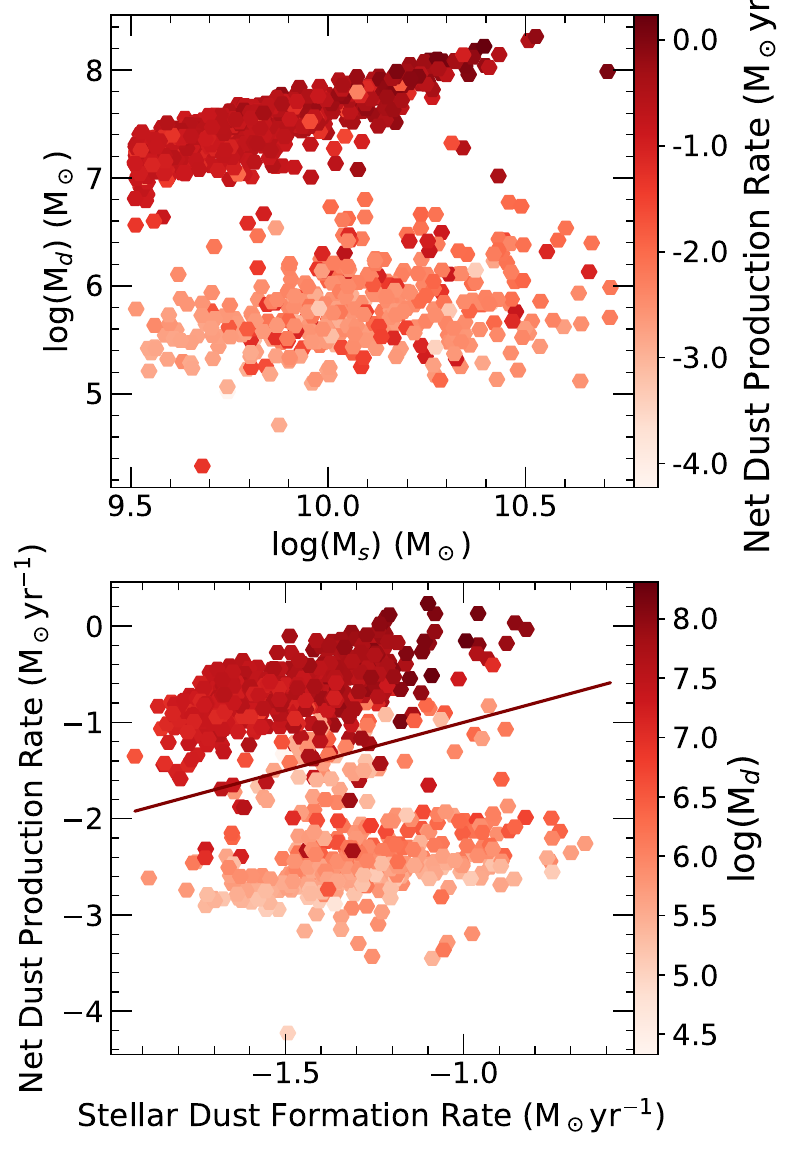}\par      
    \caption{Dust mass versus stellar mass, color-coded by the net dust production rate (top panel), and the net dust production rate versus dust formation rate by AGB stars, SNe~Ia, and SNe~II (bottom panel), color-coded by dust mass. The solid line represents the one-to-one relation. The net dust production rate is defined as the difference between the total formation rate (stellar production plus ISM growth) and the SN destruction rate.}
\vspace{-0.3cm}
\label{rates}
\end{figure}

In this section we present our results -- divided into three parts. The first examines how well our model reproduces the dust scaling relations at the redshifts of the ACJ galaxies (i.e.\ $z \sim 4.5$--5.7). The second follows the dust enrichment history of these galaxies and investigates the relative contributions of different dust formation channels. The final part explores the environments of ACJ-like galaxies in the model.

Selecting model galaxies in exactly the same way as in the observations is not straightforward, given the selection criteria summarised in the previous section. We therefore adopt a selection based on the physical properties of the observed galaxies. Specifically, we select model galaxies according to their redshift ($z \sim 4.5$--5.7), stellar mass ($M_\star \sim 10^{9.5}$--$10^{10.9}\,{\rm M}_\odot$), star formation rate (SFR $\sim 10^{1.2}$--$10^{2.7}\,{\rm M}_\odot\,{\rm yr}^{-1}$), and metallicity ($12+\log({\rm O/H}) \sim 7.8$--8.7), consistent with the values measured for the ACJ sample (\citealp{Faisst25}). This selection yields a sample of 33845 model galaxies, of which 1213 host dust reservoirs for the MS run. We tested the robustness of our results against variations in the selection criteria and find that our conclusions remain unchanged (see Appendix~\ref{appC}). The following results are based on the model run using the MS merger trees, while corresponding results based on the PMS merger trees are presented in Appendix~\ref{appB}. The PMS simulation has a mass resolution roughly one order of magnitude higher than MS, and provides a higher time resolution, particularly at high redshifts. We examined predictions using a sub-volume of the PMS to test the robustness of our conclusions. Our conclusions remain unchanged.

\subsection{Dust properties}

Figure \ref{Z-SFR-Md} shows the gas-phase metallicity (left), SFR (middle), and dust mass (right) as functions of stellar mass in the redshift interval $4.5 \leq z \leq 5.7$. The red hexbins show the two-dimensional density distribution of the model predictions. Cyan and gray stars correspond to the ACJ and ALPINE samples, respectively. Cyan error bars indicate the typical uncertainties of the ACJ measurements. In the right panel, cyan stars with black centres represent detections, while the ones without centres represent non-detecations (upper limits). Filled (CEERS; \citealp{Burgarella25}) and open (REBELS; \citealp{Algera25}) circles represent measurements of normal star-forming galaxies at $z \sim 4.5$ -- 6. Filled (\citealp{Dudzeviciute20}) and open (\citealp{daCunha15}) squares represent measurements of submillimeter galaxies at $z \sim 4$ -- 6.  

The left and middle panels reflect the selection criteria adopted in this work, expressed in the form of scaling relations between the corresponding physical quantities. Although the selected model galaxies are constrained to match the observed values of the individual physical quantities, this does not automatically ensure that the model reproduces the corresponding scaling relations. Nevertheless, the model agrees reasonably well with both observed relations.

In contrast, we did not impose any further selection in constructing the right panel; therefore, this relation constitutes a genuine prediction of the model. Two model galaxy populations emerge: dust-rich and dust-poor systems. The dust mass predictions for the dust-rich population are in good agreement with the dust measurements of all the observational datasets shown.\footnote{Conservative uncertainties on dust masses amount to $\sim 0.3$ dex (see, e.g., the discussion in O26). For some of the observational samples shown here, the uncertainties can reach $\sim 0.6$ dex.} The dust mass measurements of these galaxy samples are broadly consistent with each other, although the submillimeter galaxies exhibit systematically higher dust masses at fixed stellar mass. These observational samples cluster almost entirely around the dust-rich model population, with the exception of one galaxy from the CEERS sample. 

\begin{figure*}
\centering
   \makebox[\textwidth][c]{\includegraphics[width=1.\textwidth]{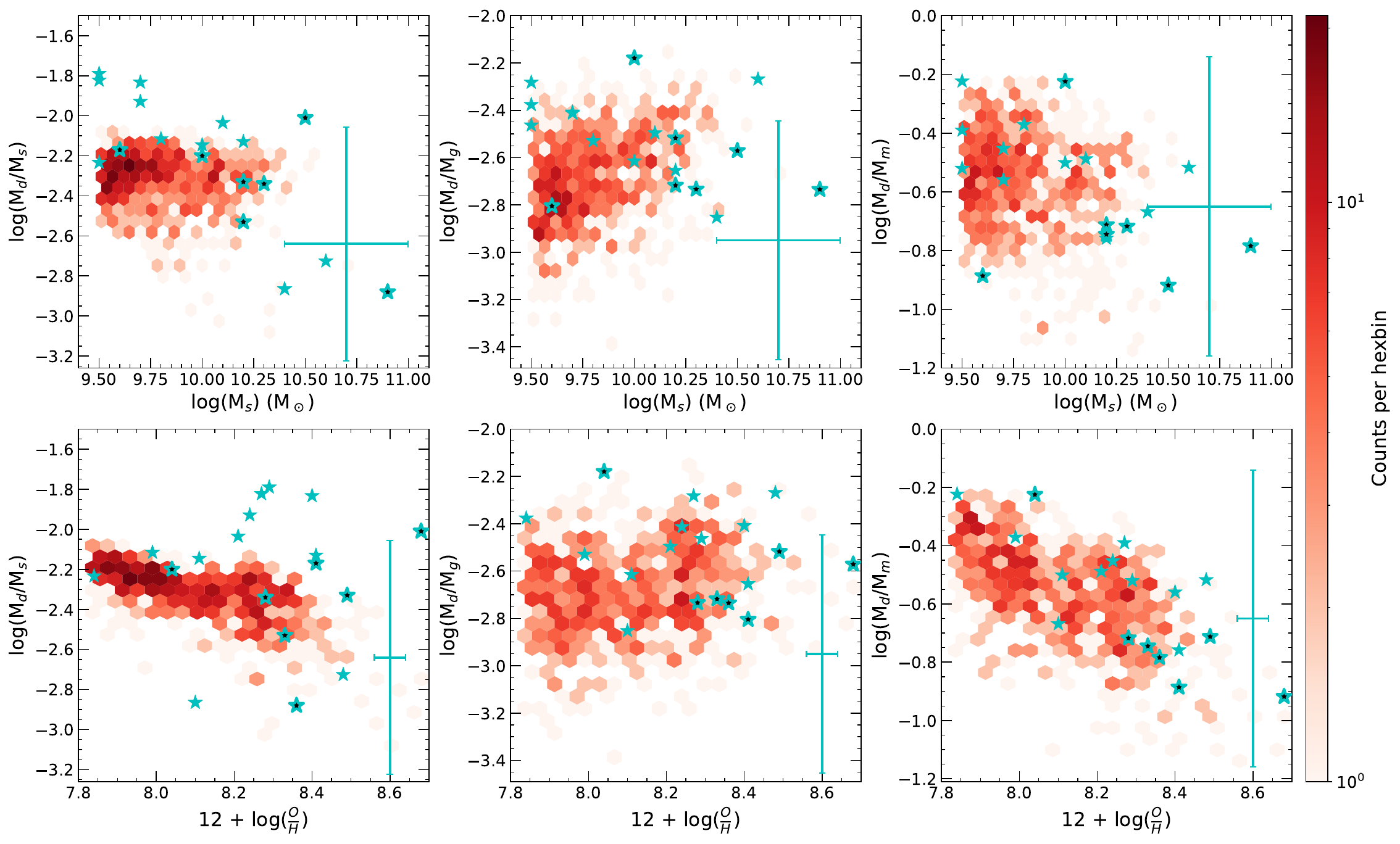}}\par    
    \caption{Dust-to-stellar mass, dust-to-gas, and dust-to-metal ratios as functions of stellar mass (top panels) and metallicity (bottom panels), from right to left, respectively. The red hexbins, cyan stars, and error bars have the same meaning as in Fig. \ref{Z-SFR-Md}.}
\vspace{-0.3cm}
\label{dust_ratios}
\end{figure*}

The dust-poor population, despite having stellar masses, metallicities, and SFRs comparable to those of the dust-rich galaxies, exhibits significantly lower dust masses and is not observed in the current datasets. To understand the origin of this dichotomy, we analysed their dust formation and destruction rates. Fig. \ref{rates} shows the dust mass as a function of stellar mass, color-coded by the net dust production rate (top panel), and the net dust production rate as a function of the stellar dust formation rate from AGB stars, SNe~Ia, and SNe~II (bottom panel), color-coded by dust mass. The solid line represents the one-to-one relation. The net dust production rate is defined as the difference between the total formation rate (stellar production plus ISM growth) and the SN destruction rate.

Dust-poor galaxies exhibit significantly lower net production rates than dust-rich systems (top panel), indicating that destruction operates more efficiently in these galaxies. Dust growth is weak or absent in these galaxies, that are therefore unable to replenish their dust reservoirs through this channel. This behavior is also reflected in the bottom panel. Assuming that destruction primarily affects dust formed through ISM growth, galaxies lying on the one-to-one line are in a steady state, where growth exactly balances destruction and the dust produced by stars remains unaffected. Galaxies below the line experience net destruction, as growth is insufficient to compensate for SN-driven destruction and even stellar-produced dust is progressively removed. Conversely, galaxies above the line are in a regime where growth exceeds destruction, leading to a net increase in dust mass.

Interestingly, we find that all galaxies in the dust-poor population experienced minor mergers with satellites of stellar mass between $10^{7}$ and $10^{8}\,{\rm M}_\odot$ in the previous time step. Their dust masses are consistent with being largely supplied by these mergers, suggesting that their dust content is primarily accreted rather than produced in situ.

The ALPINE--CRISTAL--JWST sample is likely incomplete at dust masses below $10^{7}\,{\rm M}_\odot$. Most of the observational dust-mass constraints are upper limits ($\sim 60\%$), and these limits are generally above $10^{7}\,{\rm M}_\odot$, indicating that the low-dust-mass population cannot be robustly assessed with the current data. We therefore restrict our following analysis to galaxies with $M_{\rm dust} > 10^{7}\,{\rm M}_\odot$ in order to compare the model only within the regime that is observationally accessible. A more appropriate cut would likely be mass dependent, since observed galaxies at the low-stellar-mass end have dust measurements below this threshold and are still reproduced by our model. However, the number of galaxies affected in this case is small.

Reproducing the dust mass--stellar mass relation alone is not sufficient for constraining dust formation channels. For example, \cite{Popping17} and O26 showed that dust evolution models remain highly degenerate when only this constraint is considered. In contrast, dust-to-gas, dust-to-metal and dust-to-stellar mass ratios provide stronger constraints on these channels, as they encode information about galaxy star formation histories, stellar dust production, destruction and growth efficiencies, grain size distributions, and chemical compositions (e.g. \citealp{Hirashita99}; \citealp{Remy14}; \citealp{Relano18}; \citealp{Umehata25}; \citealp{Algera25}). We therefore examine these ratios as functions of both stellar mass and metallicity.

Figure \ref{dust_ratios} shows the dust-to-stellar mass, dust-to-gas, and dust-to-metal ratios as functions of stellar mass (top panels) and metallicity (bottom panels). The red hexbins, cyan stars, and error bars have the same meaning as in Fig. \ref{Z-SFR-Md}. Starting with the relations as a function of stellar mass (top panels), most of the model galaxies have stellar masses below $10^{10.4}$ M$_\odot$, slightly lower than the upper range covered by the observational data. All three relations show a relatively small intrinsic scatter, with a 1$\sigma$ dispersion below $\sim$0.3 dex. Using a Spearman rank correlation test on our modeled galaxies, we find that the strongest correlation is for the dust-to-gas ratio as a function of stellar mass ($r = 0.346$, $p < 10^{-5}$), while the dust-to-stellar mass and dust-to-metal ratios show weak anti-correlations ($r = -0.10$ and $-0.14$, respectively, with $p < 0.05$).

Moving to the relations as a function of metallicity (bottom panels), the model spans a metallicity range that is consistent with that covered by the observational data (by construction). The scatter in all relations is reduced compared to the stellar mass trends, with a 1$\sigma$ dispersion below $\sim$0.2 dex. In this case, the predicted dust-to-stellar mass and dust-to-metal ratios show strong anti-correlations with metallicity ($r = -0.65$ and $-0.58$, respectively, with $p < 10^{-5}$), while the dust-to-gas ratio exhibits a weak positive correlation ($r = 0.158$, $p < 10^{-5}$). 

Overall, the model reproduces the observed dust scaling relations as functions of both stellar mass and metallicity reasonably well. The ACJ data points lie within the main distribution of model galaxies in all panels, indicating that the model captures both the normalization and the scatter of the observed relations. A slight offset is visible in the dust-to-stellar mass ratio, where the model tends to underestimate the values, particularly when compared to upper limits derived from non-detections at fixed stellar mass and metallicity. At the highest stellar masses, where the number of model galaxies becomes sparse, the observational data follow a natural extension of the model trends.

\subsection{Enrichment history}

\begin{figure*}
\centering
   \makebox[\textwidth][c]{\includegraphics[width=1.1\textwidth]{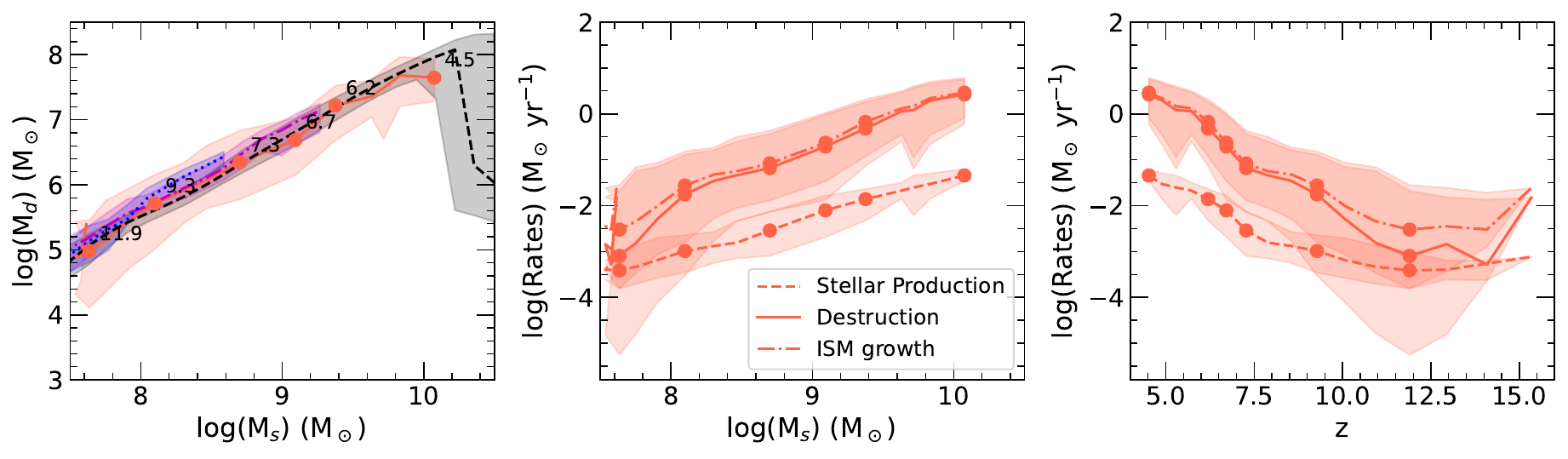}}\par    

    \caption{Dust mass versus stellar mass (left panel) and dust formation and destruction rates versus stellar mass (middle panel) and redshift (right panel). Shaded areas represent the 16th to 84th percentile region. In the left panel, redshift is indicated along the track. Points along the tracks in the middle and right panels correspond to the same redshift values as in the left panel. In the middle and right panels, the dashed, solid, and dash-dotted lines represent the formation by stars, destruction, and ISM growth rates, respectively. For comparison, the left panel also shows the predictions for the full galaxy population at $z\sim4.5$, 8, and 11, indicated by the black dashed, magenta dash-dotted, and blue dotted lines, respectively.}
\vspace{-0.3cm}
\label{z_evolution}
\end{figure*}

Fig.~\ref{z_evolution} shows the build-up of dust mass as a function of stellar mass (left panel), and the dust formation and destruction rates as functions of stellar mass (middle panel) and redshift (right panel). The shaded regions indicate the 16th--84th percentile range. In the left panel, the redshift corresponding to each point along the evolutionary track is indicated. For clarity, these labels are not repeated in the middle and right panels, where the points correspond to the same redshift values. In these panels, the dashed, solid, and dash-dotted lines show the stellar dust formation, dust destruction, and dust growth rates, respectively. For comparison, the left panel also shows the predictions for the full galaxy population at $z\sim4.5$, 8, and 11, indicated by the black dashed, magenta dash-dotted, and blue dotted lines, respectively.

The left panel shows that ACJ-like galaxies experience rapid dust mass build-up, consistent with their fast stellar mass growth and metal enrichment (\citealp{Faisst25}). The dust mass shows a strong correlation with stellar mass ($r \sim 1$, $p  < 10^{-5}$), with a slope of $\sim 1.12$, indicating that dust builds up nearly in lockstep with stellar mass in these systems. By redshift $z \sim 6.7$, these galaxies have already assembled stellar masses of $\sim 10^9$ M$_\odot$ and dust masses of $\sim 10^{6.6}$ M$_\odot$, about 0.4 dex higher than galaxies with similar stellar masses in the local Universe (e.g. \citealp{Remy14}; \citealp{Grossi15}; \citealp{DeVis19}). Over a relatively short timescale ($\sim$ 530 Myrs, from $z \sim 6.7$ to $z \sim 4.5$), they increase both their stellar and dust masses by approximately one order of magnitude on average. The predictions for the full galaxy population show negligible evolution of the dust mass--stellar mass relation between $z\sim11$ and 4.5. The evolutionary track of ACJ-like galaxies closely follows these relations at the corresponding redshifts, indicating that their dust enrichment at fixed stellar mass is broadly representative of the overall galaxy population. Thus, although ACJ-like galaxies undergo rapid dust build-up, they do not appear to be preferentially dust enriched relative to galaxies of similar stellar mass throughout their evolutionary histories. At $z\sim4.5$, the relation for the full galaxy population shows a sharp decline at the high-mass end due to the onset of black hole (of mass $>$ 10$^7$ M$_\odot$) feedback, which removes dust from the ISM through outflows and locks part of it into the growing black hole.

The middle and right panels show that, for most of their evolution, dust growth in the ISM and destruction rates are considerably higher than the formation rates by stars. The growth rate is typically slightly higher than the destruction rate, allowing the net production rate (formation + growth - destruction) to remain positive at all redshifts. The middle panel further shows that while the stellar dust formation rate increases steadily with stellar mass as the redshift decreases, both growth and destruction rates drop significantly at the low-mass end above $z \sim 8.5$. The right panel highlights a similar behaviour in redshift space: at high redshift, all rates evolve slowly, while they increase more rapidly toward lower redshifts. Similar trends are found in a smaller sample of ACJ-like galaxies drawn from the run based on the Planck Millennium merger trees (219 dusty galaxies; see Appendix \ref{appB}), where the evolution can be traced to higher redshifts and lower stellar masses. In this case, the rates remain nearly constant down to $z \sim 9$, and then increase rapidly toward $z \sim 4.5$. Together, these trends indicate a transition in the efficiency of dust production processes around $z \sim 8-9$.

\subsection{Environment}

Approximately 50\% of the ACJ sample is classified as merger pairs (\citealp{Romano21}) or likely merging/interacting systems (\citealp{Posses25}; \citealp{Faisst26}), indicating that these galaxies reside in environments where close encounters are common and structure growth is actively ongoing (\citealp{Romano21}; \citealp{Ginolfi20}). This is broadly consistent with our model, in which approximately 52\% of the dust-rich ACJ-like galaxies experience at least one merger between $z\sim4.9$ and 4.5.\footnote{Unlike the dust-poor galaxy population, mergers of these galaxies bring in a small fraction of the total dust reservoir.} Motivated by these observational and theoretical indications, we investigate the environments of ACJ-like galaxies in our model. In the following analysis, we present the results separately for the three redshift bins, distinguished by different colours.\footnote{The same redshift bins are used throughout the paper and are chosen to match the redshift distribution of the ACJ sample.} 

\begin{figure}
    \includegraphics[height=12cm,width=8cm]{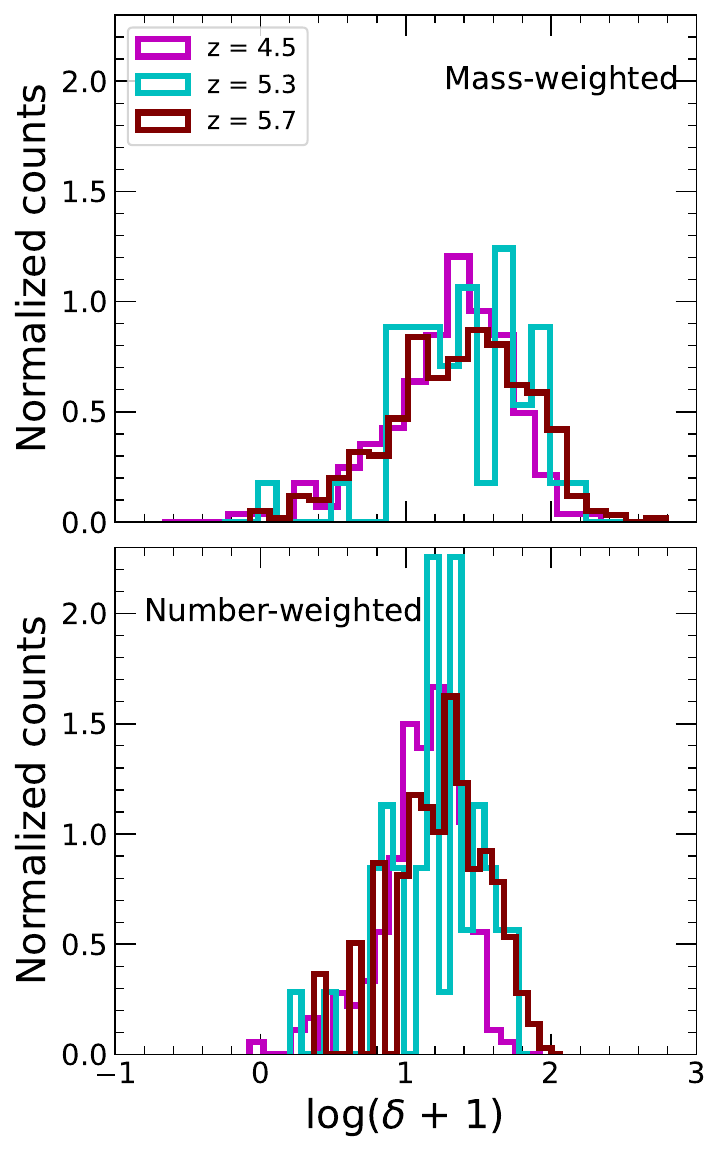}\par      
    \caption{Normalized distributions of the mass-weighted (top) and number-weighted (bottom) overdensity around dusty ACJ-like galaxies as described in the text. The coloured histograms correspond to different redshift bins.}
\vspace{-0.3cm}
\label{OvD}
\end{figure}

Following \citet{DeLucia26}, we estimated the overdensity around each dusty ACJ-like galaxy. This was done by measuring the mean overdensity of galaxies within a sphere of radius 3 Mpc/$h$, normalized by the mean overdensity obtained at each redshift from 10,000 randomly drawn spheres of the same radius within the full simulation volume. We included all galaxies with stellar mass above 10$^{8.5}$ M$_\odot$ in our calculation. We also repeated the analysis using radii of 5 and 10 Mpc/$h$, obtaining similar results. We estimated both mass-weighted and number-weighted overdensities. When calculating the mass-weighted and number-weighted overdensities, we use the total stellar mass of galaxies enclosed within the sphere and the total number of galaxies, respectively. In both estimates, we exclude the the central ACJ-like galaxies. Figure \ref{OvD} shows the normalized distributions of these quantities in the top and bottom panels, respectively, with coloured histograms corresponding to the different redshift bins.

Both overdensity definitions indicate that ACJ-like galaxies reside in a wide range of environments. At z $\sim$ 4.5, the mass-weighted overdensity has a median of 1.33 dex, with a 16th--84th percentile range of 0.87--1.69 dex, while the number-weighted overdensity has a median of 1.15 dex and a 16th--84th percentile range of 0.88--1.37 dex. At z $\sim$ 5.7, these values increase slightly to a median mass-weighted overdensity of 1.39 dex, with a 16th--84th percentile range of 0.93--1.84 dex, and a median number-weighted overdensity of 1.27 dex, with a 16th--84th percentile range of 0.97--1.57 dex. To quantify the evolution between the lowest and highest redshift bins, we performed two-sample Kolmogorov--Smirnov tests and calculated the Wasserstein distances. For the mass-weighted distribution, we obtain D = 0.117, p = 0.051, and W = 0.077 dex, indicating no statistically significant difference at the 5\% level and only a small distributional shift. In contrast, for the number-weighted distribution, we obtain D = 0.241, p = 3.6 $\times$ 10$^{-7}$, and W = 0.132 dex, indicating a statistically significant difference, with the z $\sim$ 5.7 sample shifted toward higher relative overdensities. Normalising the overdensity by the average density of the Universe (the simulation volume in this case) at each redshift ensures that this shift is not a mere consequence of the fact that the Universe is denser at higher redshifts. 

Overall, these results suggest that ACJ-like galaxies inhabit broadly similar mass-weighted environments over the redshift range considered, while the number-weighted overdensity distributions show stronger evidence for mild evolution toward denser relative environments at earlier cosmic times.

\begin{figure*}
\centering
   \makebox[\textwidth][c]{\includegraphics[width=1.1\textwidth]{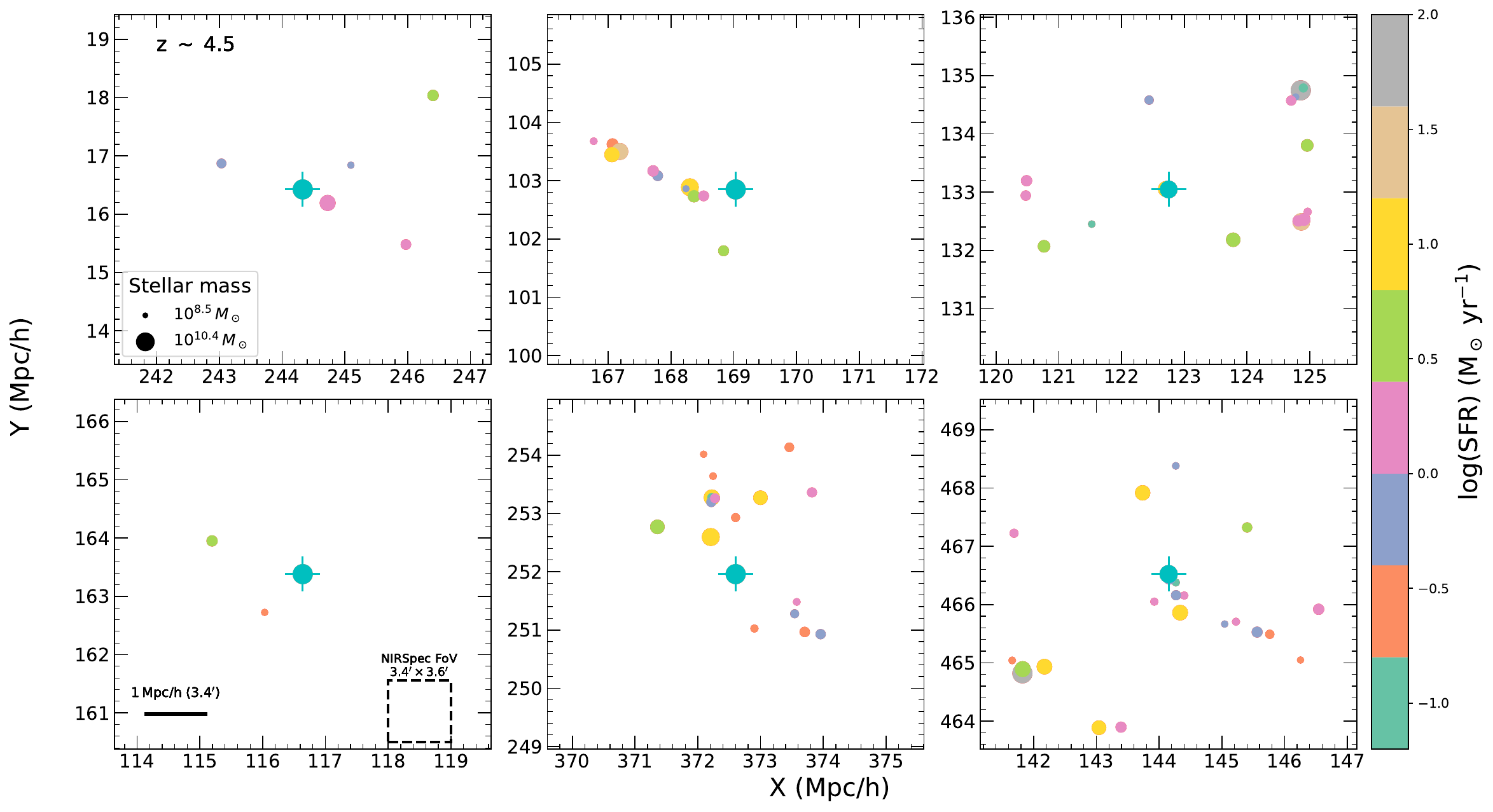}}\par    
    \caption{Projected $x$–$y$ positions of neighbouring galaxies within a sphere of radius 3 Mpc/$h$ around ACJ-like galaxies at $z \sim 4.5$. The first, second and last columns illustrate two random examples of the environment around ACJ-like galaxies (cyan dots) drawn from the lowest 16 percentiles, the 1$\sigma$ range around the median, and the highest 84 percentiles, respectively,  of the mass-weighted overdensity distribution measured as described in the text. The sizes of the circles scale with stellar mass, including the ACJ-like galaxies, with larger circles representing more massive galaxies, while their colours indicate the SFR. For reference, the dashed square in the bottom-left panel shows the NIRSpec field of view at this redshift.}
\vspace{-0.3cm}
\label{2D}
\end{figure*}

Figure~\ref{2D} shows the projected $x$–$y$ positions of neighbouring galaxies within a sphere of radius 3 Mpc/$h$ around ACJ-like galaxies at $z \sim 4.5$. Each column presents two randomly selected ACJ-like galaxies (cyan circles) drawn from the lowest 16th percentile, the 1$\sigma$ range around the median, and the highest 84th percentile of the mass-weighted overdensity distribution, from left to right, respectively. The sizes of the circles, including those representing the ACJ-like galaxies, scale with stellar mass, with larger circles corresponding to more massive galaxies, while their colours indicate the SFR. For reference, the dashed square in the bottom-left panel shows the NIRSpec field of view at this redshift. The figure illustrates the wide variety of environments inhabited by ACJ-like galaxies, ranging from systems with only a few nearby neighbours to considerably richer environments. Several of the neighbouring galaxies are relatively massive and actively star-forming, and could therefore be detectable in observations targeting the surroundings of ACJ-like galaxies (e.g. the yellow circles).

\section{Discussion}
\label{Discussion}

\subsection{Relative role of different dust formation channels}

The selection criteria adopted in this work were originally introduced to study the metal enrichment history of the ACJ sample (\citealp{Faisst25}), where we argued that ACJ galaxies have likely experienced rapid chemical enrichment. In this work, we use the same selection to investigate their dust enrichment history, without imposing any additional selection on dust mass. We note that there are well-known strong correlations between dust and stellar mass, SFR, and metallicity (\citealp{daCunha10}; \citealp{Casey12}; \citealp{Corbelli12}; \citealp{Santini14}; \citealp{Beeston18}; \citealp{Nersesian19}; \citealp{Rowlands14}; \citealp{Remy14}; \citealp{Popping22}). Stellar mass and SFR trace the overall level of dust enrichment, while metallicity regulates the relative importance of the different dust formation channels, particularly the onset and efficiency of dust growth in the ISM. For example, \citet{Santini14}, using a sample of galaxies out to $z \sim 2.5$, showed that much of the scatter in the dust mass--stellar mass relation can be explained by variations in SFR at fixed stellar mass.

Selecting sources based on stellar mass, SFR, and metallicity, we find our model galaxies reproduce reasonably well the overall dust enrichment level, as traced by the dust mass--stellar mass relation (Fig. \ref{Z-SFR-Md}), as well as the dust-to-stellar mass, dust-to-gas, and dust-to-metal ratios (Fig. \ref{dust_ratios}). These additional constraints are essential for distinguishing between dust evolution models that would otherwise remain degenerate when considering only the dust mass--stellar mass relation (\citealp{Popping17}; O26). This is because each ratio probes different aspects of dust evolution. The dust-to-stellar mass ratio primarily traces the integrated efficiency of stellar dust production, while the dust-to-gas and dust-to-metal ratios are more sensitive to the efficiencies of dust growth and destruction. In addition, these ratios are expected to depend on the grain size distribution and dust composition (e.g. \citealp{Kuo12}; \citealp{Asano14}). The ability of the model to simultaneously reproduce all these observational constraints (Figs. \ref{Z-SFR-Md} and \ref{dust_ratios}) provides confidence that it captures the dominant processes governing dust enrichment and can be used to investigate the enrichment history of ACJ galaxies. 

\begin{figure*}
\centering
   \makebox[\textwidth][c]{\includegraphics[width=1.1\textwidth]{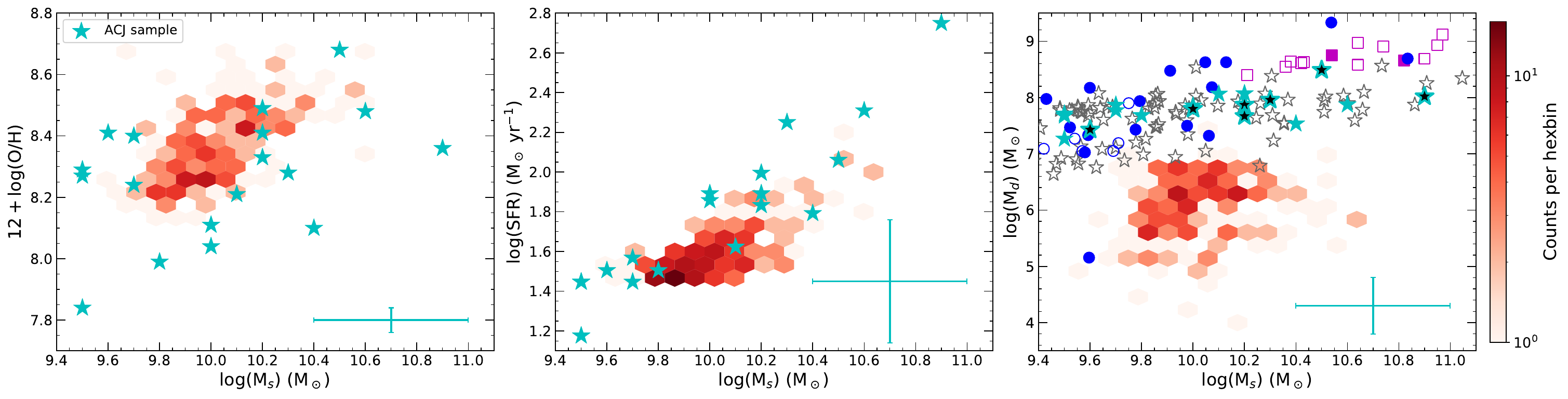}}\par    
    \caption{Same as Fig. \ref{Z-SFR-Md} for the model without dust growth.}
\vspace{-0.3cm}
\label{SFR-Z-Md-Ms_no}
\end{figure*}

Our results indicate that dust growth in the ISM is required to reproduce these observational data. To test the robustness of this conclusion, we consider a model variant in which dust growth is disabled while adopting the maximum possible dust condensation efficiencies in stellar ejecta, i.e. $\delta^{\rm SNII}=1$, $\delta^{\rm AGB}=1$, and $\delta^{\rm SNIa}=0.4$. Figure~\ref{SFR-Z-Md-Ms_no} shows the predictions of this model variant.\footnote{To construct this figure, we used a representative volume of the MS run.} Even under these extreme assumptions, the model significantly underestimates the dust masses of the ACJ galaxies. This reflects the high efficiency of dust destruction by SNe in these massive, actively star-forming systems. As the SN rate increases with stellar mass, stellar dust production alone is unable to compensate for dust destruction, requiring an additional contribution from dust growth to build up the observed dust reservoirs at these redshifts. In the absence of dust growth, dust masses comparable to, or slightly exceeding, the observed values are reached only when dust destruction is also switched off.

Observational studies of galaxies at even higher redshifts ($z \sim 7$ -- 8) have shown that dust growth is a viable mechanism for explaining the observed dust enrichment levels (e.g. \citealp{Umehata25}; \citealp{Algera25}). Beyond $z \sim 8$, however, detecting dust emission in the submillimetre becomes increasingly challenging, and dust is inferred primarily from the UV continuum slope. Using deep ALMA and NOEMA observations, \citet{Bakx26} placed an upper limit of $\sim10^{5}$ M$_\odot$ on the average dust mass of galaxies beyond $z \sim 8$. Taken together, current observations suggest a dichotomy in dust properties across these cosmic epochs, with substantial dust reservoirs commonly observed below $z \sim 7$ -- 8, while little or no dust inferred at earlier times (e.g. \citealp{Arrabal23}; \citealp{Morales24}; \citealp{Topping24}). From a theoretical perspective, this behaviour may be linked to the chemical enrichment level of galaxies. Toward higher redshifts, the ISM metal content decreases, increasing the dust growth timescale and reducing the relative importance of growth compared to stellar dust production. Following the enrichment history of ACJ-like galaxies, we find that a transition occurs around $z \sim 8$ -- 9, slightly offset with respect to results from \cite{Narayanan26}, who found a similar transition at $z \sim 10$ -- 11 using zoom-in hydrodynamical simulations, but consistent with results from \cite{Burgarella26}, who find a transition around $z \sim 8.9$.

\begin{figure}
\centering
   \includegraphics[height=8.cm,width=8.cm]{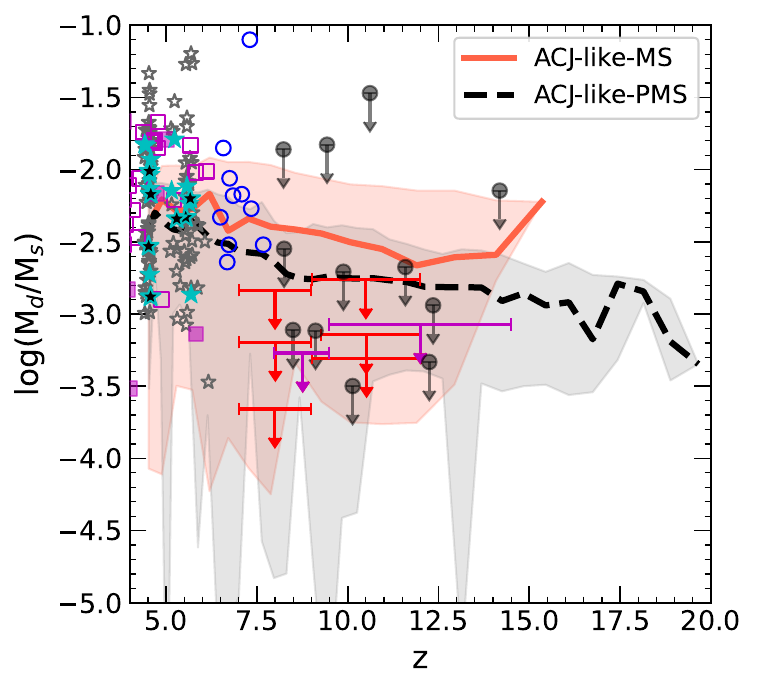}\par    
    \caption{Dust-to-stellar mass ratio as a function of redshift. The red solid and black dashed lines show the model predictions from the MS and PMS runs, respectively, with the shaded regions indicating the corresponding $3\sigma$ spread. Cyan stars with black centres represent detections in the ACJ sample, while cyan stars without black centres indicate non-detections (upper limits). Open circles show measurements for the REBELS sample (\citealp{Algera25}), while filled (\citealp{Dudzeviciute20}) and open (\citealp{daCunha15}) squares show measurements for submillimetre galaxies. The black upper limits and the red and magenta stacked measurements show the estimates adopted by \citet{Bakx26}.}
\vspace{-0.3cm}
\label{DMs}
\end{figure}

The progenitors of ACJ-like galaxies may indeed be difficult to detect in dust emission at these redshifts. Fig.~\ref{DMs} compares the evolution of their dust-to-stellar mass ratios with available observational constraints. The red solid and black dashed lines show the model predictions from the MS and PMS runs, respectively, with the shaded regions indicating the corresponding $3\sigma$ spread. Cyan stars with black centres represent detections in the ACJ sample, while cyan stars without black centres indicate non-detections (upper limits). Open circles show measurements for the REBELS sample (\citealp{Algera25}), while filled (\citealp{Dudzeviciute20}) and open (\citealp{daCunha15}) squares show measurements for submillimetre galaxies. The black upper limits and the red and magenta stacked measurements show the estimates adopted by \citet{Bakx26}.

The median predictions of both the MS and PMS runs remain above the stacked measurements and most of the upper limits at $z\gtrsim8$. However, the predicted $3\sigma$ ranges are broad and encompass both the stacked measurements and the upper limits, extending from $\log(M_{\rm d}/M_\star)\sim-2$ to values below $-4$. This comparison suggests that galaxies at $z\gtrsim8$ with dust properties similar to those predicted for the progenitors of ACJ-like galaxies may be only marginally detectable, or remain undetected, in current observations. Moreover, the lower end of the predicted distribution shows that a subset of these progenitors can reach $\log(M_{\rm d}/M_\star)<-3$, placing them firmly within the regime currently dominated by non-detections.

\subsection{Dust-rich galaxies as signposts of large over-densities}

The high fraction of ACJ galaxies classified as merger pairs (\citealp{Romano21}) or likely merging/interacting systems (9/18; \citealp{Posses25}; \cite{Faisst26}) suggests that these galaxies reside in environments where close encounters are more common and structure growth is actively ongoing (\citealp{Romano21}; \citealp{Ginolfi20}). In addition, three galaxies in the sample (DC-842313, DC-848185, and DC-873321) have serendipitously detected [C\,II]-emitting companions (\citealp{Loiacono21}). Several of the serendipitously detected [C\,II] emitters may also belong to known or candidate proto-cluster structures: in particular, DC-848185 is a known member of the AzTEC-3 proto-cluster at $z\sim5.3$, while DC-842313 may lie in the outskirts of the PCI~J1001+0220 proto-cluster at $z\sim4.5$. \cite{Loiacono21} therefore argued, for their full sample of 12, that the serendipitous [C\,II] emitters may trace both small-scale interacting systems and larger-scale overdense environments. Two more galaxies of the sample, VC-5101218326 and VC-5110377875, are confirmed spectroscopic members of PCI~J1001+0220 (\citealp{Lemaux18}). Taken together, these results suggest that at least a subset of ACJ galaxies inhabit locally overdense regions.

Our model predictions are consistent with this picture, indicating that ACJ-like galaxies reside in environments that are typically $\sim 20$ times denser than the cosmic mean, or $\sim 14$ times denser when using the unweighted estimate (Fig. \ref{OvD}), with little redshift evolution over the interval considered. Galaxies in dense environments are expected to experience evolutionary processes earlier or more efficiently than their counterparts in average-density regions (reviews by \citealp{Overzier16}; \citealp{Alberts22}). At high redshift, such environments may enhance gas accretion, increase the frequency of interactions and minor mergers, and sustain high star-formation activity. In turn, rapid star formation accelerates metal production, while the resulting dense and metal-rich interstellar medium favours efficient dust growth. In this picture, the environment may represent one of the factors contributing to the rapid build-up of stellar mass, metals, and dust in these massive galaxies in the early Universe. This would be consistent with observations that proto-clusters are often associated with excesses of dust-obscured star formation (review by \citealp{Overzier16}).

\section{Summary and conclusions}
\label{sum}
In this paper, we have used the semi-analytic galaxy formation model GAEA to study the dust enrichment history of the ACJ galaxy sample, a benchmark sample for studies of normal star-forming galaxies at redshifts z $\sim$ 4.5–6 (\citealp{Faisst25}). Our main findings are summarised below.

-- Our model reproduces the main dust-related observational constraints at the redshifts of the ACJ galaxies, namely the dust mass--stellar mass relation, and the dust-to-stellar mass, dust-to-gas, and dust-to-metal ratios as functions of stellar mass and metallicity. Reproducing these trends requires efficient dust growth in the ISM. The combination of stellar mass, SFR, and metallicity provides a good predictor of the dust properties of galaxies, as a consequence of significant correlation between these physical properties that our model reproduces well.

--  ACJ-like galaxies experience rapid dust enrichment, consistent with their rapid metal enrichment. We also find evidence for a transition redshift around z $\sim$ 8–9, below which the dominant dust production channel shifts from stellar sources to ISM dust growth. This is consistent with recent observational findings. The broad range of predicted dust-to-stellar mass ratios at earlier epochs further suggests that galaxies at $z \gtrsim 8$ with dust properties similar to those predicted for the progenitors of ACJ-like galaxies may remain undetected in current dust-continuum observations.

-- Our model is consistent with the interpretation that ACJ galaxies preferentially reside in locally overdense environments. The rapid build-up of stellar mass, metals, and dust, along with the likely dense environments of these systems, suggests that environment may have contributed to their accelerated evolution.

\begin{acknowledgements}

We are grateful to F. Pozzi, T. Bakx, and D. Zakharova for stimulating discussions and feedback, and to D. Burgarella for provideing us with tabulated information for the CEERS data. We acknowledge support from the INAF 2022 Theory Grant ``The critical role of DUST in the cosmic barYon cycle”. OO and FF acknowledge support from the INAF 2024 RSN1 Minigrant “The effect of a variable IMF on Galaxy Evolution and Assembly: from the local to the high-z Universe.”. This work has been partially supported by the Italian Ministry of University and Research (MUR) Missione 4 ``Istruzione e Ricerca” - Componente C2, Investimento 1.1 Fondo per il Programma Nazionale di Ricerca e Progetti di Rilevante Interesse Nazionale (PRIN), the PRIN 2022 PNRR grant under the National Recovery and Resilience Plan (PNRR): project P2022ZLW4T ”Next-generation computing and data technologies to probe the cosmic metal content”. We also acknowledge the use of INAF-OATs computational resources within the framework of  the INAF PLEIADI program (Taffoni et al. 2020 http://www.pleiadi.inaf.it). LX acknowledges support from the National Natural Science Foundation of China (grant number 12041302), the Ministry of Science and Technology of China (grant No. 2020SKA0110100). 
\end{acknowledgements}
 
\bibliographystyle{aa}
\bibliography{bibliography}
\begin{appendix}

\onecolumn

\section{Alternative selection of the ACJ-like galaxies}
\label{appC}

\begin{figure*}
\centering
   \makebox[\textwidth][c]{\includegraphics[width=1.1\textwidth]{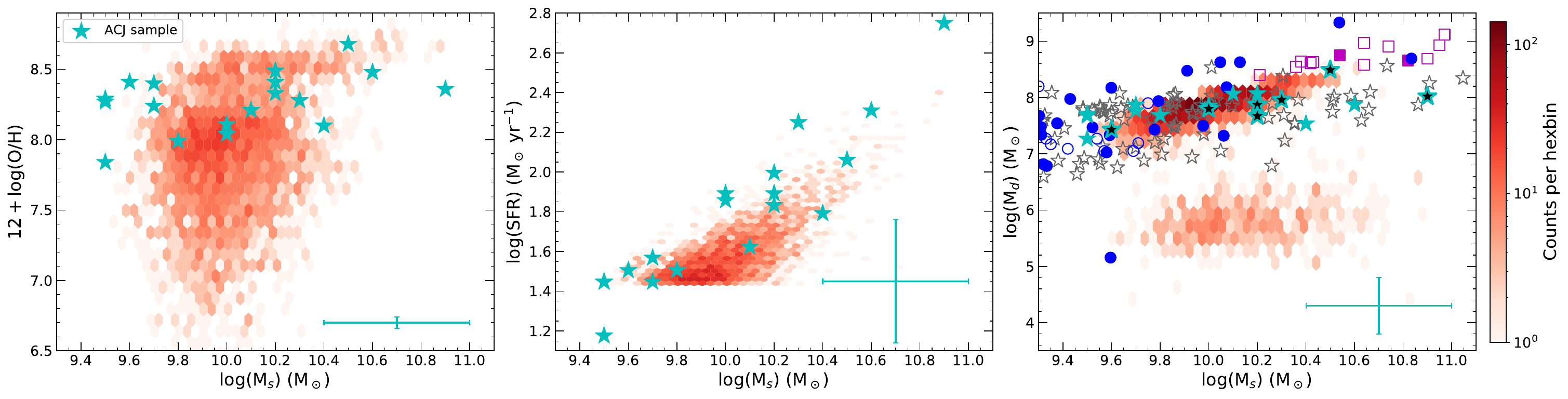}}\par    
    \caption{Same as Fig. \ref{Z-SFR-Md} for an alternative selection of the ACJ-like galaxies (see the text for further details).}
\vspace{-0.3cm}
\label{SFR-Z-Md-Ms_selection}
\end{figure*}

Due to the complexity of the ACJ sample selection, which we cannot reproduce exactly, we test an alternative selection criterion for our model galaxies representing the ACJ systems. In this test, we limit our analysis to the model galaxies at $z \sim 4.5$. In this selection, we relax the constraints on metallicity, as well as the upper limits on stellar mass and SFR. In other words, we include all galaxies with stellar masses $M_\star \geq 10^{9.5},{\rm M}\odot$ and SFR $\geq 10^{1.44},{\rm M}\odot,{\rm yr}^{-1}$. Figure~\ref{SFR-Z-Md-Ms_selection} shows the results of this selection. The figure indicates that ACJ galaxies preferentially occupy the upper envelope of the mass–metallicity relation predicted by the model. Our conclusions remain unchanged.

\section{Properties of the ACJ-like galaxies in the PM run}
\label{appB}

\begin{figure*}
\centering
   \makebox[\textwidth][c]{\includegraphics[width=1.1\textwidth]{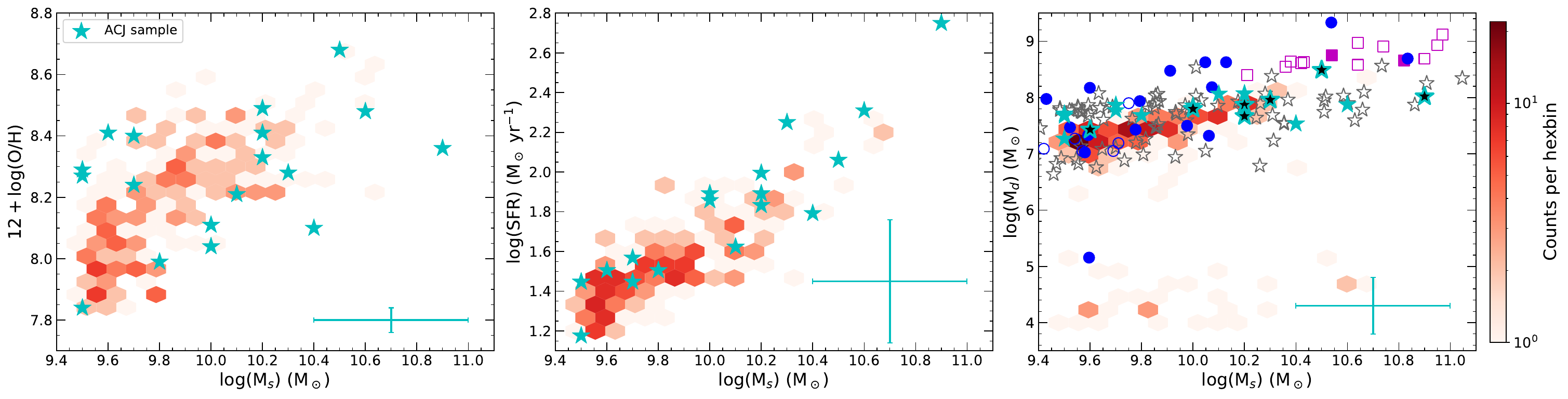}}\par    
    \caption{Same as Fig. \ref{Z-SFR-Md} for the PM model run.}
\vspace{-0.3cm}
\label{SFR-Z-Md-Ms_PM}
\end{figure*}

\begin{figure*}
\centering
   \makebox[\textwidth][c]{\includegraphics[width=1.1\textwidth]{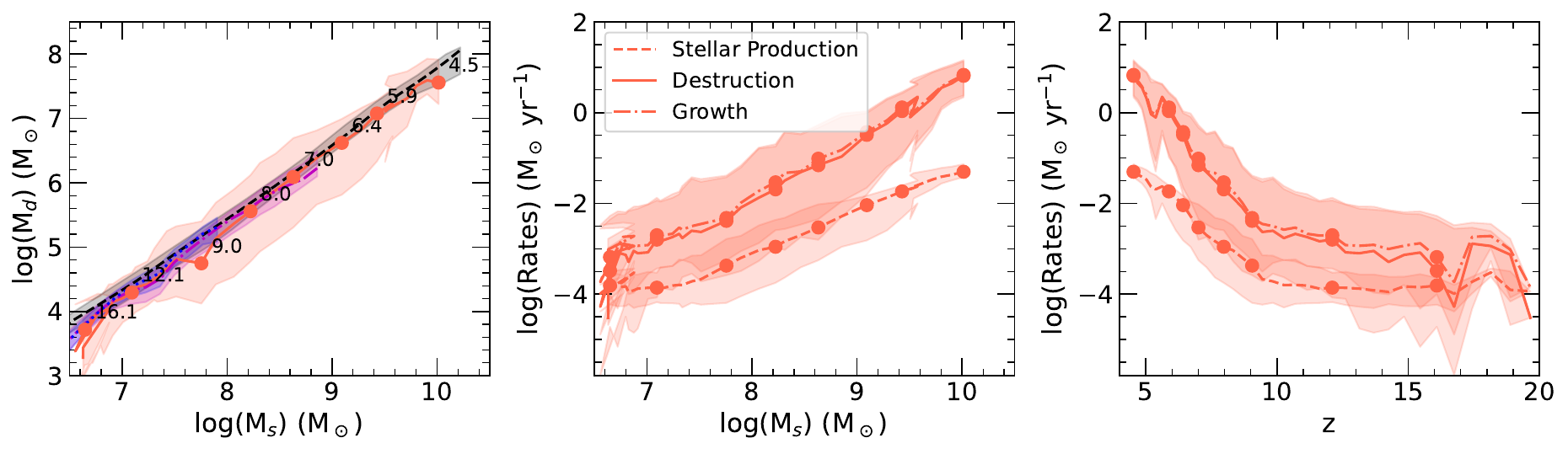}}\par    
    \caption{Same as Fig. \ref{z_evolution} for the PM model run.}
\vspace{-0.3cm}
\label{z_evolution_PM}
\end{figure*}

We repeat our analysis using the model run based on the PM merger trees (see Section~\ref{sec:sec2} for details of the simulation). We limit our analysis to a subvolume in which ACJ-like galaxies have a high occurrence rate. The resulting sample contains 2355 ACJ-like galaxies, of which 219 have large dust reservoirs. Figures~\ref{SFR-Z-Md-Ms_PM} and \ref{z_evolution_PM}, analogous to Figs.~\ref{Z-SFR-Md} and \ref{z_evolution}, respectively, show the results of this analysis. Overall, our conclusions remain unchanged.

\section{Additional galaxy properties}
\label{appA}

\begin{figure*}
\centering
   \makebox[\textwidth][c]{\includegraphics[width=1.1\textwidth]{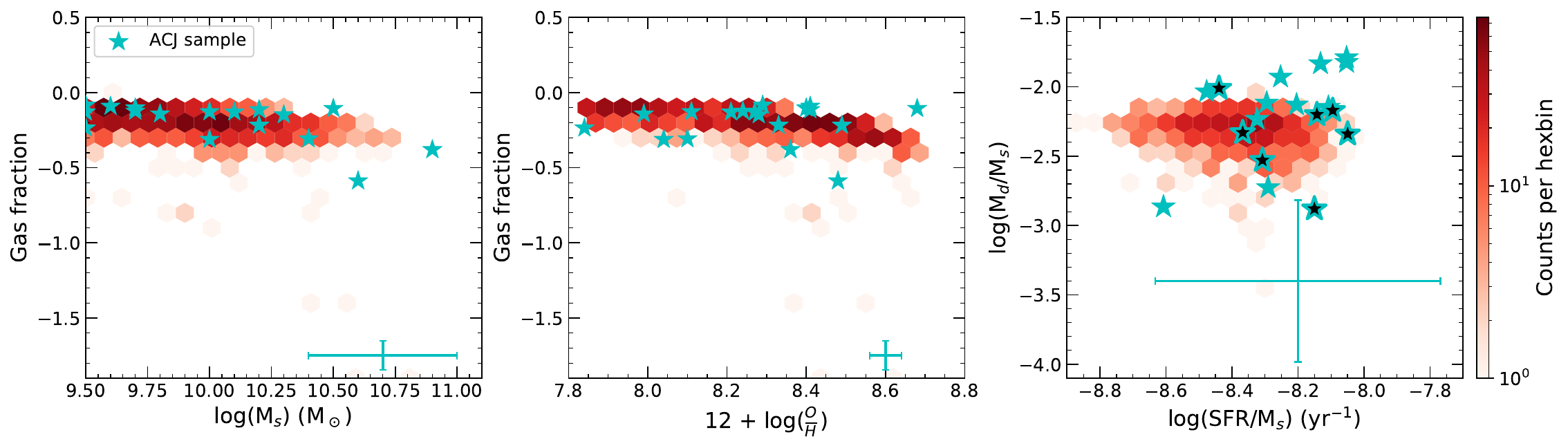}}\par    
    \caption{Gas mass (left) and gas mass fraction (right) as functions of the stellar mass of the ACJ-like galaxies using the MS run. The cyan stars and error bars, as well as the red hexbins, have the same meaning as in Fig.~\ref{Z-SFR-Md}.}
\vspace{-0.3cm}
\label{FG}
\end{figure*}

In this appendix, we present additional galaxy properties, namely the gas fraction [$M_g/(M_g + M_\star)$, where $M_g$ and $M_\star$ are the gas and stellar masses, respectively] as a function of stellar mass and metallicity, and the specific dust mass as a function of specific star formation rate, shown in the left, middle, and right panels of Fig.~\ref{FG}, respectively, using the MS run. The cyan stars and error bars, as well as the red hexbins, have the same meaning as in Fig.~\ref{Z-SFR-Md}. The model reproduces all three scaling relations reasonably well.

\end{appendix}

\end{document}